\documentclass[aps,prl,showpacs,twocolumn,floatfix,superscriptaddress]{revtex4}
\usepackage{color}
\usepackage{bm}
\usepackage{amssymb}
\usepackage{amsmath}
\usepackage{amstext}
\usepackage{graphics}
\usepackage{epsfig}
\usepackage{placeins}
\usepackage{psfrag}
\usepackage{subfigure}

\begin{document}

\title{Supersolid Order of Frustrated Hard-Core Bosons in a Triangular
Lattice System}
\author{H. C. Jiang}
\affiliation{Center for Advanced Study, Tsinghua University, Beijing, 100084, China}
\author{M. Q. Weng}
\affiliation{Department of Physics, University of Science and Technology of China, Hefei
230026, China}
\author{Z. Y. Weng}
\affiliation{Center for Advanced Study, Tsinghua University, Beijing, 100084, China}
\author{D. N. Sheng}
\affiliation{Department of Physics and Astronomy, California State University,
Northridge, California 91330, USA}
\author{L.~Balents}
\affiliation{Department of Physics, University of California, Santa Barbara, California
93106}
\date{\today }

\begin{abstract}
We numerically demonstrate that a supersolid phase exists in a  frustrated
hard-core boson system on a triangular lattice over a wide  range of
interaction strength. In the infinite repulsion (Ising)  limit, we establish
a mapping to the same problem with unfrustrated  hopping, which connects the
supersolid to the known results in that  case. The weak superfluidity can be
destroyed or strongly  enhanced by a next nearest neighbor hopping term,
which provides  valuable information for experimental realization of a
supersolid phase  on optical lattice.

\end{abstract}

\maketitle

\textit{Introduction:} A supersolid phase is a state of matter
exhibiting both diagonal and off-diagonal long-range order (ODLRO)
\cite{andreev}. Recent possible observation of a supersolid phase
\cite{he} in $^{4}$He under pressure has attracted a lot of
interest. While the microscopic conditions under which clean
$^{4}$He could be in a supersolid phase are still unclear,
supersolidity is established for hard-core bosons on a triangular
lattice, which is the focus of many recent
studies\cite{moessner,qmc1,qmc2,qmc3,variation}. The supersolid
phase is an example of ordering by disorder, demonstrated for
hard-core boson system with unfrustrated nearest neighbor (NN)
hopping and strong repulsion\cite{qmc1,qmc2,qmc3} based on extensive
Quantum Monte Carlo (QMC) simulations and theoretical analysis.
These theoretical works are motivated in part by experimental
realizations of lattice bosons in ultra-cold atom
traps\cite{optical}. Intriguingly, the superfluid density in the
supersolid phase is found to be very small, possibly indicating that
the system is near a phase boundary\cite{burkov} to an insulating
phase. It is thus highly desirable to examine the stability of the
supersolid phase in more extended models to determine the relevant
perturbation and possibly to suggest a route of getting into a deep
supersolid phase for experiment.

The nature of the state for the hard-core bosons with
\emph{frustrated} NN hopping on triangular lattice is another open
issue, where the model can be mapped to the spin-1/2 XXZ
antiferromagetic (AF) Heisenberg model which suffers from the sign
problem. Historically this model was the first candidate proposed to
realize a spin liquid ground state \cite{anderson}, although it
turns out to still exhibit a three sub-lattice AF long-range-order
(LRO) in general, which may persist to large $J_{z}$
limit\cite{read}. However, extensive numerical studies have been
limited to near the SU(2) point\cite{bernu}, and the precise nature
of the ordering at larger $J_{z}$ (or the strong NN repulsion limit
for the corresponding boson model) has not been well understood.

In this Letter, we present a systematic density-matrix
renormalization group (DMRG) and exact diagonalization (ED)
numerical studies of the half-filled ground state of the
\emph{frustrated} model over a wide range of the NN repulsion. We
show that a robust supersolid phase does exist in this model, and it
can be related to the well-known supersolid phase of an unfrustrated
NN hopping model in infinite repulsion (Ising) limit, where a
precise mapping between the two models by a sign transformation can
be established. Furthermore, we reveal that the supersolid phase is
close to a transition to an insulating phase, and correspondingly
the superfluidity can be strongly enhanced (or easily switched off)
by tuning a next nearest neighbor (NNN) hopping term.

\textit{Supersolid Phase from Isotropic SU(2) Point to Large $J_{z}$ Limit:}
We consider a simple model of hard-core bosons at half-filling on the
triangular lattice interacting via a repulsive term,
\begin{equation}
H=-t\sum_{\langle ij\rangle }\left( b_{i}^{\dag }b_{j}+b_{j}^{\dag
}b_{i}\right) +\sum_{\langle ij\rangle }V(n_{i}-\frac{1}{2})(n_{j}-\frac{1}{2%
}),  \label{eq:1}
\end{equation}%
where $b_{i}^{\dagger }$ is a boson creation operator and $n_{i}$ a boson
number operator. $\langle ij\rangle $ denotes NN sites and we shall mainly
consider the frustrated boson hopping at $t<0$. This boson Hamiltonian is
equivalent to, by a standard mapping from hard-core bosons to S=1/2 spins,
an XXZ Heisenberg model on the triangular lattice:
\begin{equation*}
\mathcal{H}={\sum_{\langle ij\rangle }}\left[ {\frac{J_{\perp }}{2}}%
(S_{i}^{\dagger }S_{j}^{-}+h.c.)+J_{z}S_{i}^{z}S_{j}^{z}\right] ,
\end{equation*}%
with $J_{z}=V$ and $J_{\perp }=-2t$. In the spin language, a charge density
wave (solid) order implies infinite-range correlations of the $z$-component
of spins at a nonzero wavevector, while the superfluid order is equivalent
to the in-plane ordering of the spins also at a nonzero wavevector for the
frustrated system. In the following analysis, we will work interchangeably
in terms of bosons and spin variables.

\begin{figure}[tbp]
\centerline{
    \includegraphics[height=3.3in,width=3.4in]{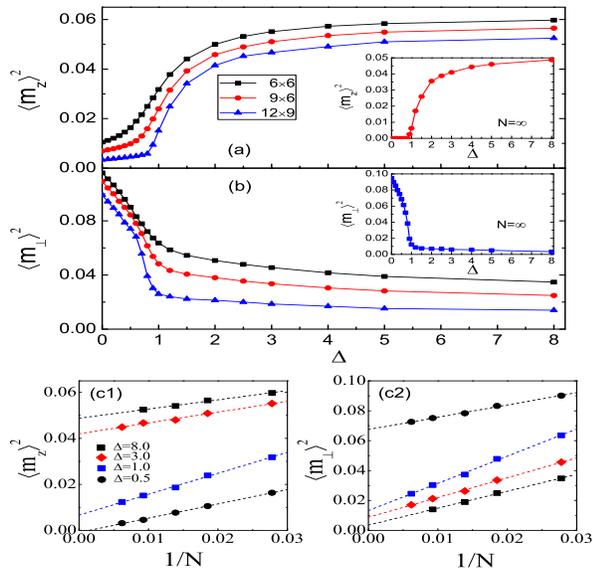}}
\caption{(color online) The order parameter squared, $\langle m_{z}\rangle
^{2}$ and $\langle m_{\perp}\rangle^{2}$, as functions of the anisotropy $%
\Delta =J_{z}/J_{\perp }$, are shown in (a) and (b) respectively, with $%
N=6\times 6$, $9\times 6$ and $12\times 9$. The insets are the
corresponding extrapolations in the thermodynamic limit. Examples
of finite-size scaling of the order parameters are also shown in
(c1) and (c2) with system size up to $N=9\times 18$.}
\label{phasediagram}
\end{figure}

We first present the numerical results based on DMRG method\cite{white} for
a triangular lattice with the total number of sites $N=N_{1}\times N_{2}$.
We keep up to $m=4096$ states in each DMRG block for most systems, and the
truncation error is of the order or less than $10^{-5}$. We make use of the
periodic boundary condition to reduce the finite-size effect for a more
reliable extrapolation to the thermodynamic limit. To analyze the magnetic
properties of the system, we calculate the structure factors $S_{z}(\mathbf{q%
})$ and $S_{\perp }(\mathbf{q})$ defined by
$S_{z}(\mathbf{q})={\frac{1}{N}}{\sum_{i,j}}e^{-i\mathbf{q}(\mathbf{r}_{i}-
\mathbf{r}_{j})}\langle S_{i}^{z}S_{j}^{z}\rangle$ and
$S_{\perp}(\mathbf{q})={\frac{1}{N}}{\sum_{i,j}}e^{-i\mathbf{q}(\mathbf{r}
_{i}-\mathbf{r}_{j})}\langle S_{i}^{\dagger }S_{j}^{-}\rangle$.

The obtained $S_{z}(\mathbf{q})$ and $S_{\perp }(\mathbf{q})$ show Bragg
peaks at the corners of the hexagonal Brillouin zone [e.g., at $q_{0}=(\pm
4\pi /3,0)$]. In particular, at small $\Delta \equiv J_{z}/J_{\perp }<1$,
the peak of $S_{z}(\mathbf{q}_{0})$ is very weak, while that of $S_{\perp }(%
\mathbf{q}_{0})$ is very sharp, representing the dominant AF correlation in
the $XY$ plane. With the increase of $\Delta $, $S_{z}(\mathbf{q_0})$ grows
continuously and its value becomes bigger than the in-plane ones passing the
point $\Delta = 1$. One can obtain the magnetic order parameters based on
the finite-size scaling of the peak values of $S_{z}(\mathbf{q}_{0})$ and $%
S_{\perp }(\mathbf{q}_{0})$. Specifically, the average magnetization $%
\langle m_{z}\rangle $ and $\langle m_{\perp }\rangle $ can be determined by
$\langle m_{z}\rangle ^{2}=S_{z}(\mathbf{q}_{0})/N$ and $\langle m_{\perp
}\rangle ^{2}=S_{\perp }(\mathbf{q}_{0})/N$, which are shown vs. $\Delta $
at $N=36$, $54$ and $108$ in the main panel of Fig. \ref{phasediagram}(a)
and (b).

Nonzero $\langle m_{z}\rangle ^{2}$ and $\langle m_{\perp }\rangle ^{2}$ in
the thermodynamic limit will correspond to the diagonal LRO and ODLRO,
respectively. Examples of the finite-size scaling are shown in Fig.\ref%
{phasediagram} (c1) and (c2) by plotting $\langle m_{z}\rangle^{2}$ and $%
\langle m_{\perp }\rangle^{2}$ as functions of $1/N$. Thus obtained order
parameters extrapolated to the thermodynamic limit are presented in the
insets of Fig. \ref{phasediagram} (a) and (b). At small $\Delta $, the
system is in the pure superfluid phase with magnetic order solely lying in
the $XY$-plane with $\langle m_{\perp }\rangle ^{2}\neq 0$ and $\langle
m_{z}\rangle =0$ (\emph{cf.} the inset of Fig. \ref{phasediagram}). By
contrast, at $\Delta >1$ the three-sublattice AFLRO develops in both $z$%
-direction and $XY$ plane, characterized by nonzero values of $\langle
m_{z}\rangle ^{2}$ and $\langle m_{\perp }\rangle ^{2}$. Here $\langle
m_{z}\rangle $ monotonically increases with $\Delta $ from the isotropic
point ($\Delta =1$), consistent with the spin-wave picture of co-planar
ordering in the $XZ$ plane. For the corresponding boson system, our results
suggest a supersolid phase with coexisting diagonal LRO and ODLRO at $\Delta
\geq \Delta _{c}$. The phase boundary $\Delta _{c}$ between the superfluid
phase and supersolid phase is very close to the isotropic point $\Delta
_{c}=(J_{z}/J_{\perp })_{c}=(V/2|t|)_{c}\sim 1.00$.

\begin{figure}[tbp]
\centerline{
    \includegraphics[height=2.0in,width=3.4in]{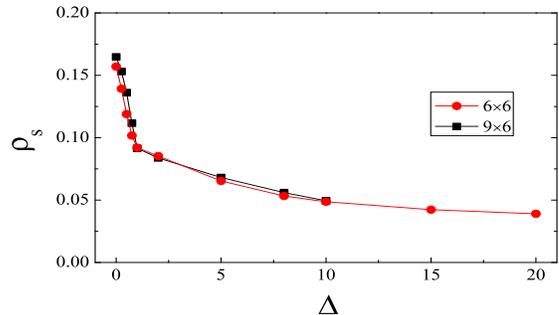}}
\caption{(color online) Superfluid stiffness $\protect\rho _{s}$ (in
units of $J_{\perp}$) as a function of the anisotropy $\Delta
=J_{z}/J_{\perp }$ obtained from ED ($N=36 $) or DMRG ($N=54$)
calculations.} \label{stiffness}
\end{figure}

We note that at large $J_{z}$ case, e.g., $\Delta =8$, $\langle m_{z}\rangle
^{2}=0.049$ ($\left\langle m_{z}\right\rangle =0.24$) which is much larger
than the corresponding value at the isotropic point, while the $XY$-plane
magnetization reduces to $\langle m_{\perp }\rangle ^{2}=0.0036$ ($%
\left\langle m_{\perp}\right\rangle =0.06$) in the thermodynamic limit.
Though this superfluid ordering is small, its value actually is comparable
with that of the unfrustrated hard-boson supersolid\cite{qmc1,qmc2,qmc3} at
the same large $J_{z}$ limit as we have checked numerically. In the
following we can further establish the presence of the superfluidity through
the calculation of the superfluid density $\rho _{s}$ by adding a nonzero
twist phase at the system boundary with $\rho _{s}=\frac{\partial ^{2}E }{%
\partial \theta _{x}^{2}}\propto \left[ E(\theta _{x}=\pi )-E(\theta _{x}=0)%
\right]$. We obtain $\rho_s$ by adding a twist phase $\theta _{x}=\pi $ in
both ED and DMRG calculations\cite{rho}, which are shown in Fig. \ref%
{stiffness} as a function of $\Delta $ for $N=36$ and $N=54$. From
the figure, we can see that the finite-size effect for $\rho _{s}$
is very weak and $\rho_s$ should remain finite in the thermodynamic
limit, consistent with the
finite $\langle m_{\perp }\rangle ^{2}$ in Fig. \ref{phasediagram}. At $%
J_{z}>10$, the DMRG becomes difficult to converge as the energy from $J_{z}$
term becomes dominant, the ED results further suggest that the obtained $%
\rho _{s}$ should be nonzero over the whole range of $J_{z}$
with a value matching with the ones for the unfrustrated bosons in the large
$J_{z}$ limit\cite{qmc3}.

\textit{Supersolid Order in the Ising Limit:} Now we turn to the
interesting limit of $\Delta \rightarrow \infty $, where the direct
in-plane magnetic ordering in the numerical results is very weak and
the SW theory suggests that it vanishes as square root of
$J_{\perp}/J_z$. Clearly, here a rigorous examination is desired. At
$J_z \rightarrow \infty$, the XXZ model in Eq.~(2) reduces to the
classical Ising AF on a triangular lattice. This classical model is
well-known to have a macroscopic degeneracy of ground states,
which correspond to all spin configurations with
exactly one frustrated bond per triangle \cite{ising}. In this
limit, the XXZ model reduces to
\begin{equation}
H_{\infty }=J_{\perp}/2\sum_{\langle ij\rangle }\hat{P}%
_{C}(S_{i}^{+}S_{j}^{-}+h.c.)\hat{P}_{C},  \label{eq:ising}
\end{equation}%
where $\hat{P}_{C}$ is a projection operator onto the classical Ising ground
state manifold (IGSM).

\begin{figure}[tbp]
\centerline{
    \includegraphics[width=2.8in]{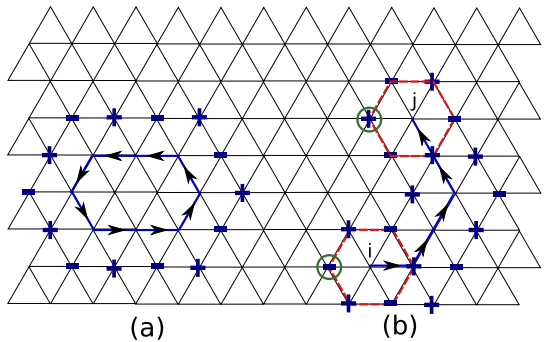}
    }
\caption{(color online) Illustration of the paths contributing to
the high temperature expansion, and the associated sign rule. In
(a), we show a representative loop appearing in the expansion of the
partition function. The $\pm$ signs show one of two alternating spin
configurations allowed around the loop, which allows only
even-length loops. In (b), we show a representative path in the
expansion of the off-diagonal correlation function between sites $i$
and $j$. For such a path with an odd number of steps, the auxiliary
sites $i+\hat{a},j+\hat{a}$ (shown in circles) must be
antiparallel.} \label{mapping}
\end{figure}

The unfrustrated model with $J_\perp<0$ has been studied previously, and
shown to exhibit supersolid order. We now show how supersolidity in the
frustrated case $J_\perp>0$ can be deduced from those known results. We
first consider thermodynamic and other properties that can be deduced from
the partition function and \textsl{diagonal} expectation values of the form
\begin{equation}  \label{eq:2}
Z[{\mathcal{O}};J_\perp] = \mathrm{Tr}\,\left[ \hat{P}_C {\mathcal{O}}[\{
S_i^z \}] e^{-\beta H_\infty}\right],
\end{equation}
where ${\mathcal{O}}$ is any function of the z-components of the spins, or
boson occupation numbers. From such quantities we can calculate the free
energy, and the diagonal (solid) correlations. We show that $Z$ is an
\textsl{even} function of $J_\perp$, and so these properties are identical
for the frustrated and unfrustrated cases. To see this, consider the high
temperature expansion of $Z$ in powers of $\beta J_\perp$. The terms in the
expansion consist of successive actions of bond operators of the form $\hat{P%
}_C S_i^+ S_j^-\hat{P}_C $ on nearest-neighbor links, with a factor of $%
\beta J_\perp$ accompanying each bond operator. To achieve a
non-zero expectation value in the trace, the boson number on each
site must be unchanged after the action of all these operators.
Graphically, we may represent each factor of $S_i^+S_j^-$ on the
lattice as an arrow pointing from site $j$ to site $i$, and we
require this ``vector field'' have zero divergence, i.e. the arrows
close into ``exchange'' loops. Now consider the contribution from
any particular state in the trace. Due to the projection, each bond
operator has a non-zero action only if $i$ and $j$ are
``flippable'', i.e. the two other spins on each triangle containing
$i$ or $j $ are anti-parallel. Now let us circumscribe each exchange
loop on our graphical representation by a neighboring loop as in
Fig.~\ref{mapping}.  In order that all sites on the exchange loops
are flippable, spins on the neighboring loops must alternate, which
requires that all of the neighboring loops must have an
\textsl{even} number of sites. This in turn requires that
the total number of links on each exchange loop is even. Thus $Z[{\mathcal{O}%
};J_\perp]$ is indeed an even function of $J_\perp$.

Now consider the off-diagonal correlation function,
\begin{equation}  \label{eq:3}
\langle S_j^+ S_i^-\rangle = Z^{-1} \mathrm{Tr}\,\left[ \hat{P}_C S_j^+
S_i^- e^{-\beta H_\infty}\right].
\end{equation}
Once again, one may consider the high temperature expansion of the numerator
(the denominator $Z$ has already been shown to be even). In this case,
contributions must be divergenceless \textsl{except} at the sites $i$ and $j$%
, which appear as source and sink, respectively. One can understand the
behavior by considering just the simplest terms, in which the arrows form a
single path connecting $i$ to $j$ (see Fig.~\ref{mapping}).
Now form a tightly circumscribing loop about this path. As above,
for any state in the trace to contribute, the spins $S^z_k$ on the
sites of this neighboring loop must alternate. Moreover, the 6 spins
neighboring $i$ and $j$ must also alternate since $S_j^+S_i^-$ acts
directly on these states. Now consider the product $4S_{i+\hat{a}}^z
S_{j+\hat{a}}^z$, where $\hat{a}$ is any nearest-neighbor vector,
acting on a state which contributes to the trace. Because of the
alternating spins around the circumscribing loop, this factor gives
the parity of the exchange path, i.e. it equals $+1$ for an even
path and $-1$ for an odd path. One may show that this conclusion is
unaffected by additional closed loops, which appear as higher order
terms in the high temperature series. Since this is true for every
term in the expansion, we
find $\langle S_j^+S_i^-\rangle|_{J_\perp>0}=\langle 4S_{i+\hat{a}}^z S_{j+%
\hat{a}}^zS_j^+S_i^-\rangle |_{-J_{\perp}}$.

\begin{figure}[tbp]
\centerline{
    \includegraphics[height=2.1in,width=3.5in]{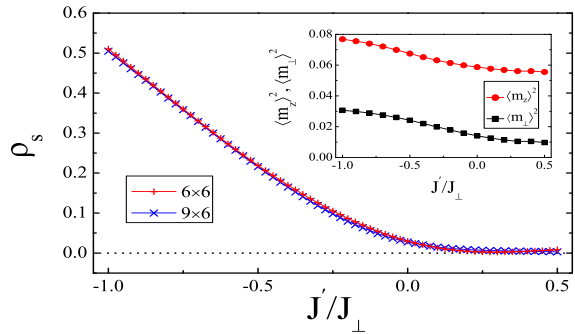}}
\caption{(color online) Superfluid stiffness $\protect\rho _{s}$ (in
units of $J_{\perp}$) vs. $J^{\prime }/J_{\perp }$ for systems with
NNN hopping $t^\prime $ (superechange $J^{\prime }$). The
finite-size order parameters $\langle m_{z}\rangle^{2}$ and $\langle
m_{\perp }\rangle^{2}$ at $N=54$ are shown in the inset.}
\label{jprime}
\end{figure}

The above observations lead us to the conclusion that the supersolid phase
survives even for the frustrated hard-core boson system at strong repulsion (%
$V=J_{z}\rightarrow \infty $) limit as it maps to the unfrustrated model\cite%
{qmc1,qmc2,qmc3}. ED calculation of the projected Hamiltonian gives the
energy per site varying between $-0.1721t$ ($N=36$) to $-0.1678t$ ($N=72$)
with a possible extrapolating value $E=-(0.162\pm 0.005)t$ in the
thermodynamic limit.

\textit{Enhancement of the Superfluidity and the Ordering of the
Supersolid Phases:} To understand the underlying reason why the
superfluid stiffness is relatively weak\cite{burkov},
we add a NNN hopping $t^{\prime }=-J^{\prime }/2$ term. For
simplicity we only present the results for $J_{\perp }=-2t>0$ and large $%
J_{z}$ (Ising) limit, although the obtained results also apply to both
models with a finite range of $J_{z}$.

The superfluid stiffness is calculated using the ED method for the
projected Hamiltonian. As illustrated in Fig. \ref{jprime},
$\rho_{s}$ is relatively small at $J^{\prime }=0$ compared to its
value in the region $J^{\prime }/J_{\perp }<0$ (negative sign
represents an unfrustrated NNN hopping). In fact, $\rho _{s}$
monotonically increases with a negative $J^{\prime }$ and when
$J^{\prime }=-J_{\perp }$, $\rho _{s}$ becomes comparable to the
value of a pure superfluid phase (i.e., the ferromagnetic XY model
on triangular lattice\cite{qmc3}). Clearly a boson system at
$J^{\prime }=0$ is indeed near the phase boundary of an insulating
phase, which occurs at $\left( J^{\prime }/J_{\perp }\right)
_{c}\sim 0.2$ (which we further identify as a solid phase with
diagonal LRO). The finite-size order parameters $\langle
m_{z}\rangle ^{2}$ and $\langle
m_{\perp }\rangle ^{2}$ for $N=54$ are also shown in the inset of the Fig. %
\ref{jprime}, where the enhancement of the peaks of the structure factors $%
S_{z}\mathbf{(}q_{0})$ and $S_{\perp }\mathbf{(}q_{0})$ are clearly seen as
we continuously turn on the negative $J^{\prime }$. Thus the resulting phase
is a supersolid phase with strong diagonal LRO and superfluidity.

These observations, and the precise nature of the supersolid ordering, can
be rationalized by simple energetic arguments in the large $J_z$ limit. For
the NN hopping case ($J^{\prime}=0$), the constraint that neighboring spins
to the hopping path must alternate tends to enhance hopping which takes ``60$%
^\circ$'' turns (forming hexagonal path), which keeps the bosons on two of
the three sublattices. Moreover, the third sublattice, on which hopping does
not proceed, must be substantially polarized. Thus the three sublattice
ordering, $\langle S_z\rangle=$($-m$, $-m$, $2m+\delta)$ is favored
energetically, consistent with a ferrimagnetic ordered phase \cite%
{variation, ashvin}. When the NNN hopping term is dominant
($-J^{\prime}\geq J_{\perp}$), bosons tend to hop on a single (say
A) sublattice, while spins on the neighboring sites from sublattices
B and C are individually preferred to be ferromagnetically aligned,
with B and C spins antiparallel to each other. This corresponds to
$\langle S_z\rangle =$($0,m,-m)$ or
``antiferromagnetic'' ordering, which we therefore expect in the large $%
-J^{\prime}$ limit. This is indeed supported by numerics, which will be
presented elsewhere.

In summary, we have established a robust supersolid phase for the
frustrated hard-core bosons on triangular lattice at half-filling
based on extensive numerical calculations and analytical analysis.
The observed supersolidity is an example of ordering by disorder
elegantly realized for such a frustrated system. Furthermore, we
have found that the supersolid phases for the hard-core boson models
with only NN hoppings are quite close to a pure solid phase in both
frustrated and unfrustrated cases. But a small unfrustrated NNN
hopping term can push the boson systems into a deep supersolid phase
with greatly enhanced superfluidity. Our theoretical study can thus
provide a solid foundation for the experimental realization of
supersolid phase on triangular optical lattice.

\textbf{Acknowledgment:} The authors would like to thank R. G. Melko
for insightful discussions. This work is supported by NSFC grant
nos. 10688401(HCJ,ZYW), 10804103 (MQW), the DOE grant
DE-FG02-06ER46305, the NSF grants DMR-0605696, DMR-0611562 (HCJ,
DNS), DMR-0804564 (LB), and the Packard Foundation (LB). In
finishing this work, we became aware of a parallel work\cite{ashvin}
which has reached the same conclusion based on different approaches.
We are grateful to the author (A.V) for sending us the preprint
before submission.

\end{document}